\documentstyle[emulateapj,psfig]{article}

\def\ltsima{$\; \buildrel < \over \sim \;$}
\def\ltsim{\lower.5ex\hbox{\ltsima}}
\def\gtsima{$\; \buildrel > \over \sim \;$}
\def\gtsim{\lower.5ex\hbox{\gtsima}}

\begin{document}


\title{AGE DATING OF A HIGH-REDSHIFT QSO B1422+231 AT $Z=3.62$ 
AND ITS COSMOLOGICAL IMPLICATIONS}

\author{Yuzuru Yoshii$^{1,3}$, Takuji Tsujimoto$^{2}$, 
and Kimiaki Kawara$^{1}$}

\altaffiltext{1}{Institute of Astronomy, Faculty of Science,
University of Tokyo, Mitaka, Tokyo 181-8588, Japan}

\altaffiltext{2}{National Astronomical Observatory, Mitaka, Tokyo 181-8588, 
Japan}

\altaffiltext{3}{Research Center for the Early Universe, 
Faculty of Science, University of Tokyo, Tokyo 113-0033, Japan}

\begin{abstract}

The observed Fe$\;${\scriptsize II}(UV+optical)/Mg$\;${\scriptsize II}
$\lambda\lambda$2796,2804 flux ratio from a gravitationally lensed
quasar B1422+231 at $z$=3.62 is interpreted in terms of detailed
modeling of photoionization and chemical enrichment in the broad-line
region (BLR) of the host galaxy.  The delayed iron enrichment by Type
Ia supernovae is used as a cosmic clock. Our standard model, which
matches the Fe$\;${\scriptsize II}/Mg$\;${\scriptsize II} ratio,
requires the age of 1.5 Gyr for B1422+231 with a lower bound of 1.3
Gyr, which exceeds the expansion age of the Einstein-de Sitter
$\Omega_0=1$ universe at a redshift of 3.62 for any value of the
Hubble constant in the currently accepted range, $H_{\rm 0}$=60--80
km\,s$^{-1}$\,Mpc$^{-1}$.  This problem of an age discrepancy at
$z=3.62$ can be unraveled in a low-density $\Omega_0\ltsim 0.2$
universe, either with or without a cosmological constant, depending on
the allowable redshift range of galaxy formation. However, whether the
cosmological constant is a required option in modern cosmology awaits a
thorough understanding of line transfer processes in the BLRs.

\end{abstract}


\section{Introduction}

Ages of distant objects at high redshift, $z>1$, unambiguously
constrain the expansion age of the universe at that redshift, and can
be used to determine the ultimate fate of the universe (Turner 1991;
Kennicutt 1996).  There is a growing interest in use of Type Ia and II
supernovae (SNe Ia and II) as a nucleosynthesis-clock in interpreting
the abundance pattern of heavy elements seen in the spectra of
high-redshift galaxies (Hamman \& Ferland 1993; Matteucci \& Padovani
1993).

Extensive calculations of explosive nucleosynthesis show that SNe II
are the sites of alpha and iron production, whereas SNe Ia produce a
large amount of iron relative to alpha elements (Nomoto, Thielemann,
\& Wheeler 1984).  Since progenitors of SNe Ia have a lifetime of
$t_{\rm Ia}\sim 1$ Gyr, longer by two or three orders of magnitude
than SNe II (Truran 1987), the switchover of the iron source from SNe
II to SNe Ia creates a break in the alpha/iron abundance ratio at a
time at which a significant number of SNe Ia start to explode.
Accordingly, the nucleosynthesis clock records such a switchover and
assigns an age above or below $t_{\rm Ia}\sim 1$ Gyr, depending on
whether the alpha/iron ratio is smaller than or equal to that of SN II
origin, respectively.  This feature acts like an alarm clock set at
$t_{\rm Ia}\sim 1$ Gyr, and motivates us to measure the alpha/iron
ratio at $z\gtsim 3$, to obtain a firm constraint on the expansion age
of the universe to be compared with that derive from the Hubble
constant, for which we adopt $H_{\rm 0}$=60--80
km\,s$^{-1}$\,Mpc$^{-1}$ (Freedman 1998) as the current best estimate.

Among various alpha elements, magnesium is ideal for comparing with
iron, because the best-studied range of rest-frame wavelengths from
1500 to 6000$\AA$ includes the emission features of Fe$\;${\scriptsize
II} multiplet lines and Mg$\;${\scriptsize
II}$\;$$\lambda\lambda$2796,2804 doublet lines, so that the ratio of
their fluxes, emergent from the same ionized zone, is a direct
indicator of the abundance ratio (Wills, Netzer, \& Wills 1985).
These wavelengths are redshifted to the near infrared at $z\gtsim 3$,
and quasars are the only objects that are bright enough for their
spectra to be studied. In this regard, Kawara et al.'s (1996) infrared
observation of the Fe$\;${\scriptsize II} to Mg$\;${\scriptsize II}
flux ratio for the gravitationally lensed quasar B1422+231 at $z=3.62$
is particularly important.  In this Letter, we compare their data with
detailed models of chemical enrichment, and attempt to determine the
age of the broad-line region (BLR) of B1422+231, which
allows us to set a combined constraint on the cosmological parameters
and the epoch of galaxy formation.

\section{Basic Equations for Chemical Evolution}

The strength of Fe$\;${\scriptsize II} emission from quasars at
various redshifts implies a much more significant iron supply from SNe
Ia to the gas in the BLR of quasars as compared to the solar
neighborhood (Wills et al. 1985; Elston, Thompson, \& Hill 1994;
Kawara et al. 1996).  Such an iron-rich gas results if quasar host
galaxies are associated with an initial burst of star formation and
then the ejected iron from SNe Ia is restored to the gas after the
burst and the cessation of star formation. This situation is explored
by modeling the chemical enrichment with two basic ingredients
(Tinsley 1980) such as (1) the star formation rate (SFR) proportional
to some power of the gas fraction $C(t)=\nu_k [f_g(t)]^k $, and (2)
the initial stellar mass function (IMF) having a time-invariant mass
spectrum $\phi(m)dm\propto m^{-x}dm$ normalized to unity between the
lower and upper mass limits $(m_l,m_u)$.

The SFR coefficient for quasar host galaxies has been constrained in a
narrow range from a variety of their observed features, that is,
$\nu_{k=1}=7.5$ Gyr$^{-1}$ from their metal abundances (Padovani \&
Matteucci 1993; Matteucci \& Padovani 1993) and $\nu_{k=1}=6.7-7.6$
Gyr$^{-1}$ from the N$\;${\scriptsize V}/C$\;${\scriptsize IV} and
N$\;${\scriptsize V}/He$\;${\scriptsize II} line ratios of
high-redshift quasars (Hamann \& Ferland 1992, 1993) having high
metallicities like B1422+231. We, therefore, adopt the higher value of
$\nu_{k=1}=7.6$ Gyr$^{-1}$ as the standard value in this paper.
Theoretical arguments indicate that the IMF originates from
fragmentation of the gas cloud which occurs almost independently of
local physics in the gas (Low \& Lynden-Bell 1976; Silk 1977).  A
solar-neighborhood IMF would therefore be a good approximation, and we
adopt a Salpeter slope of $x=1.35$ and a mass range from $m_l=0.05$ to
$m_u=50M_\odot$ (Tsujimoto et al. 1997). Then the gas fraction
$f_g(t)$ and the heavy-element abundance $Z_i(t)$ in the gas change
with time according to
\begin{equation}
\frac{df_g}{dt}=-C(t)+
\int^{m_u}_{{\rm max}(m_l,m_t)} \hspace{-1cm} dm\phi(m)r(m)C(t-t_m) 
\;\;\; ,
\end{equation}
and
\begin{eqnarray}
\frac{d(Z_if_g)}{dt}=&-&Z_i(t)C(t) \nonumber \\
&+&
\int^{m_u}_{{\rm max}(m_l,m_t)} \hspace{-1cm} dm A \phi(m) y_{{\rm Ia},i}
\int^t_0dt_{\rm Ia} g(t_{\rm Ia})C(t-t_{\rm Ia}) \nonumber \\
&+&\int^{m_u}_{max(m_l,m_t)} \hspace{-1cm} dm (1-A) \phi(m)
\{y_{{\rm II},i} \nonumber \\
&&\hspace{1cm}+Z_i(t-t_m)r_w(m)\}C(t-t_m) \;\; ,
\end{eqnarray}
respectively, where $m_t$ is the turnoff mass when the main-sequence
lifetime, $t_m$, is equal to time $t$, $r(m)$ the fraction of the
ejected material from a star of mass $m$, $r_w(m)$ the fraction of the
ejected material without newly synthesized elements from that star,
and $y_i$ the heavy-element yield from an SN II or Ia.  Since all
these stellar quantities, either calibrated or constrained by nearby
stars, should also apply to any galaxy, we can use the formula of
Renzini \& Buzzoni (1986) for $m_t$ and the updated nucleosynthesis
calculations by Nomoto's group (Tsujimoto et al. 1995) for
high-redshift galaxies.

The fraction of stars that eventually produce SNe Ia is $A=0.055$ for
$3-8M_\odot$ and $A=0$ outside this mass range.  The lifetime of their
progenitors is $t_{\rm Ia}\approx 1.5$ Gyr and its possible spread is
modeled using the power-law distribution function $g(t_{\rm
Ia})\propto t_{\rm Ia}^{\gamma}$ ($\gamma\geq 0$) which is bounded in
a specified range of $t_{\rm Ia}$ and normalized to unity. These basic
quantities $A$ and $t_{\rm Ia}$ for SNe Ia have been constrained
mainly from the observed [O/Fe] break at [Fe/H]$\sim -1$ in the solar
neighborhood (Yoshii, Tsujimoto, \& Nomoto 1996). From theoretical
considerations Kobayashi et al. (1998) recently proposed that SNe Ia
should occur only when the progenitor's metallicity is above a
critical value of [Fe/H]$\sim -1$.  While this implies that the [O/Fe]
break at [Fe/H]$\sim -1$ may only weakly constrain $t_{\rm Ia}$, the
introduction of a critical metallicity for SNe Ia resolves an
observational puzzle (Carney 1996): The Galactic globular
clusters (GCs), spanning an age difference of several Gyrs, exhibit a
constant alpha/iron abundance ratio of the SN II origin all the way
from [Fe/H]$\sim -2$ to $-0.5$ without a break.  In fact, with this
critical metallicity, the GC data place even firmer limits of $t_{\rm
Ia}=1-3$ Gyrs from the box-shaped $g$-distribution with
$\gamma=0$ (Tsujimoto \& Yoshii 1998).  We therefore use this constraint
on $t_{\rm Ia}$ to date the delayed enrichment of iron relative to
alpha elements in high-redshift galaxies.

\section{Age Dating for the B1422+231 System}

The rest-frame UV--optical spectrum of B1422+231 was obtained with the
IR detector array mounted on the KPNO 4m telescope by Kawara et
al. (1996).  The observed Fe$\;${\scriptsize
II}(UV+optical)/Mg$\;${\scriptsize II}$\;\lambda\lambda$2796,2804 flux
ratio is $12.2\pm3.9$, {\footnote{The synthetic spectrum with a
$\lambda^{-1}$-type extinction law fitted to the data gives a dust
extinction below $A_V=0.1$ mag for B1422+231, so that this effect
hardly changes our measurement of the Fe$\;${\scriptsize
II}/Mg$\;${\scriptsize II} flux ratio.}  } which is comparable to 8.9
derived from the composite spectrum of intermediate-redshift quasars
at $z=1-2$ (the Large Bright Quasar Survey (LBQS), Francis et
al. 1991) and $7.8\pm2.6$ for nine low-redshift quasars at
$z=0.15-0.63$ (Wills et al. 1985).  Using photoionization models where
important excitation processes are all included, Wills et al. (1985)
showed that within the wide range of physical parameters in their
solar-abundance models the Fe$\;${\scriptsize II}/Mg$\;${\scriptsize
II} flux ratio is firmly bounded between 1.5 and 4 with a typical
value of 3 for the likely optical depth $\tau$(Bac)$=0.5-1$ of the
hydrogen Balmer continuum. Therefore, the observed strength of
Fe$\;${\scriptsize II} emission from their sample yields an
overabundance of Fe by a factor of 3 with respect to Mg (see also
Netzer \& Wills 1983; Collin-Souffrin, Hameury, \& Joly 1988; Graham).
Accordingly, with a reasonable assumption of Fe$\;${\scriptsize
II}/Mg$\;${\scriptsize II} $\propto$ Fe/Mg, the flux ratio for
B1422+231 leads to an abundance ratio of
[Mg/Fe]$=-0.61^{+0.12\;+0.16}_{-0.30\;-0.12}$
([Mg/Fe]$\equiv\log$(Mg/Fe)$-\log$(Mg/Fe)$_{\odot}$), where the first
errors quoted correspond to the uncertainties in transforming the flux
to abundance ratio and the second to the observational errors in
measuring the flux ratio.  Throughout this paper we use the
transformation formula $[I($Fe$\;${\scriptsize
II}$)/I($Mg$\;${\scriptsize II}$)]_\odot=3^{+1}_{-1.5}$.  We note that
its uncertainty, especially in the direction of giving a larger
[Mg/Fe] (or smaller age) of B1422+231 is easily exceeded by the
combined uncertainties of the flux measurement and chemical evolution
model which will be discussed below (see Table 1).

Figure 1 shows the Fe$\;${\scriptsize II}/Mg$\;${\scriptsize II} flux 
ratio (upper panel) and the logarithmic iron abundance [Fe/H] (lower panel) 
as a function of time in units of Gyrs.  Using the Salpeter IMF and the 
nucleosynthesis prescriptions as described above, we calculate the standard 
model 
($\nu_{k=1}=7.6$ Gyr$^{-1}$; thick line) together with the lower SFR model 
($\nu_{k=1}=6.7$ Gyr$^{-1}$; dashed line). 
The Fe$\;${\scriptsize II}/Mg$\;${\scriptsize II} flux ratio is
initially maintained at a low level reflecting the low Fe/Mg abundance
ratio in the SN II ejecta, and then starts to increase with the
enhanced Fe supply due to the onset of SNe Ia from 1 Gyr until 3 Gyr.
Thereafter, the Fe$\;${\scriptsize II}/Mg$\;${\scriptsize II} flux
ratio declines, because the metal-deficient gas having the low Fe/Mg
abundance ratio of genuine SN II origin is released in the
interstellar matter from the surface of low-mass stars, with lifetimes
greater than 3 Gyr, that were born in the initial burst of star
formation. The horizontal line in the upper panel is the observed
Fe$\;${\scriptsize II}/Mg$\;${\scriptsize II} flux ratio for
B1422+231, and the shaded region brackets the range of the errors.

The intersection between the theoretical curve by the standard model
and the horizontal line representing the observed Fe$\;${\scriptsize
II}/Mg$\;${\scriptsize II} flux ratio in the upper panel gives the age
of 1.50 Gyr for B1422+231, which, according the lower panel, requires
[Fe/H]=+0.96 at that age. This high metallicity of ten times solar
agrees well with the result derived by Hamann \& Ferland (1992, 1993)
from the N$\;${\scriptsize V}/C$\;${\scriptsize IV} and
N$\;${\scriptsize V}/He$\;${\scriptsize II} line ratios of
high-redshift QSOs with $z=2-4$.

We have examined the uncertainty of the estimated age and metallicity
by repeating the calculations with different values of input
parameters from our standard choice.  Table 1 tabulates $\Delta
t\equiv t$(chd)$-t$(std) for each of the changed parameters and shows
that the standard model already uses the parameter values which result
in a relatively low estimate of the age.  Use of a higher $A$(SN Ia)
in the model makes this age smaller by $\sim$ 0.1 Gyr, thus imposing a
lower bound of 1.3 Gyr, if the uncertainties of the observed
Fe$\;${\scriptsize II}/Mg$\;${\scriptsize II} flux ratio is taken into
account. This result can be applied unless $A$(SN Ia) in the BLRs is
drastically larger than the value constrained in the solar neighborhood.

\section{Comparison with the Expansion Age of the Universe}

Consider a galaxy at redshift $z$, which is assumed to form at
$z_F$. The age of this galaxy is the time taken by the universe to
expand from $z_F$ to $z$.
Because of its high redshift of $z=3.62$, the lower age bound of 1.3
Gyr for B1422+231 constrains not only the cosmological parameters but
also the epoch of galaxy formation in a highly correlated manner.
Figure 2 shows the ($\Omega_0, H_0$) region allowed by the existence
of B1422+231.  Two values of $z_F=7$ and 10 are assumed for either a
universe with $\lambda_0=0$ (thin lines) or a flat universe with
$\Omega_0+\lambda_0=1$ (thick lines).  Evidently, the Einstein-de
Sitter universe ($\Omega_0=1$, $\lambda_0=0$) requires $H_0\ltsim 40$
km\,s$^{-1}$\,Mpc$^{-1}$ and $z_F\gtsim 10$.  However, the currently
accepted range for $H_0$ does not allow a value as small as $40$
km\,s$^{-1}$\,Mpc$^{-1}$ for $H_0$.

Figure 3 shows the predicted age--redshift relations for three 
representative combinations of the cosmological parameters: 
the Einstein-de Sitter universe ($\Omega_0=1$, $\lambda_0=0$; left panel), 
an open universe ($\Omega_0=0.2$, $\lambda_0=0$; middle panel), and 
a flat, $\Lambda$-dominated universe ($\Omega_0=0.2$, $\lambda_0=0.8$; 
right panel).  Solid lines in each panel denote the result for $H_0$= 60 
(upper line) and 80 (lower line) km\,s$^{-1}$\,Mpc$^{-1}$, assuming 
$z_F=10$.  For the purpose of comparison, the result for $z_F=5$ and
$H_0$= 60 km\,s$^{-1}$\,Mpc$^{-1}$ is shown by the dashed line. 

Figures 2 and 3 indicate that the age discrepancy, now encountered at
$z=3.62$ in the Einstein-de Sitter universe, is alleviated either in
an open, $\Omega_0\ltsim 0.2$ universe with $z_F\gtsim 10$, or in a
flat, $\Omega_0\ltsim 0.2$ universe with $z_F\gtsim 6-7$. With an age
set at the lower bound of 1.3 Gyr, we obtain two solutions for two
different ranges of $z_F$. However, with an age of 1.5 Gyr as obtained
with our standard model, an open universe becomes difficult unless
$\Omega_0\ltsim 0.1$ and $z_F\gtsim 10$.

We note that the IMF in the star-bursting phase is considered to have
a shallower slope (Contini, Davoust, \& Consid\`{e}re 1995) and/or a
larger $m_l$ (Rieke et al. 1993) than in the quiescent solar
neighborhood. Such an IMF weighted towards massive stars was also derived
from the N/C ratio in the BLRs (Hamman \& Ferland 1992, 1993).  If this were
the case in B1422+231, a longer age, exceeding even 1.5 Gyr, is
required to reach the observed Fe$\;${\scriptsize
II}/Mg$\;${\scriptsize II} flux ratio (see Table 1).  The existence of
such an old galaxy at $z=3.62$ would rule out an open universe, and
nonzero-$\Lambda$ would have to be invoked. More Fe$\;${\scriptsize
II}/Mg$\;${\scriptsize II} observations in high-redshift QSOs are
evidently needed.

The interesting possibility of a requirement for nonzero-$\Lambda$ has
recently received some support from new distance determinations to SNe
Ia discovered near $z=1$ (Perlmutter et al. 1998; Garnavich et
al. 1998).

\section{Conclusion}

Detailed models of chemical enrichment are used to estimate the age of
a gravitationally lensed quasar B1422+231 ($z=3.62$) from the
Fe$\;${\scriptsize II}/Mg$\;${\scriptsize II} flux ratio.  Our
standard model, which matches the data, requires an age of 1.5 Gyr
with a lower bound of 1.3 Gyr, whereas the Einstein-de Sitter universe
($\Omega_0=1$, $\lambda_0=0$) is 0.6--0.8 Gyr old at that redshift
with current estimates of the Hubble constant $H_0=60-80$
km\,s$^{-1}$\,Mpc$^{-1}$. This problem of the age discrepancy, now
encountered at $z=3.62$, was first identified at $z\approx 0$ by
dating the GCs (15.8$\pm$2.1 Gyr, Bolte \& Hogan 1995; 14.56$\pm$2.49
Gyr, Chaboyer et al. 1996) and the radioactive element $^{232}$Th in a
very old halo star (CS22892-052) ($15.2\pm3.7$ Gyr, Cowan et
al. 1997), and then at $z=1.55$ by dating the UV-optical spectral
energy distribution for a high-redshift radio galaxy LBDS 53W091
($\gtsim 3.5$ Gyr, Spinrad et al. 1997).  

In finding potential ways to remove the age discrepancy of the GCs,
a brighter RR Lyare calibration of $M_V$(RR)$\approx +0.2$ mag, and
therefore younger GC ages, was inferred from Hipparcos measurements of
parallaxes of Galactic Cepheids (Feast \& Catchpole 1997) and Galactic
metal-poor subdwarfs (Reid 1997).  However, an analysis of Hipparcos
proper motions and parallaxes of Galactic RR Lyraes by Tsujimoto,
Miyamoto, \& Yoshii (1998) {\it directly} confirmed $M_V$(RR)$\approx
+0.6$ mag, as previously accepted. Therefore, the age
discrepancy of the GCs is not yet resolved.

The age discrepancy at $z=3.62$ in this paper is based on the Fe/Mg
abundance ratio derived from a simple scaling of their emission line
fluxes and therefore needs to be confirmed by more detailed modeling
of the emission line transfer of Fe$\;${\scriptsize II} in the BLR
using the newly available atomic data (e.g., Sigut \& Pradhan 1998).
In this regard, although the present analysis already suggests a
nonzero-$\Lambda$ universe (see Fig. 3), a definitive discrimination
between an open universe and a nonzero-$\Lambda$ universe awaits future
efforts of elaborating the photoionization model as well as observing
the Fe$\;${\scriptsize II}/Mg$\;${\scriptsize II} flux ratio in more
QSOs at $z\sim 3-5$, because no other age-dating methods except for
the nucleosynthesis-clock are able to be employed at such a great distance
and early epoch in the universe.

\acknowledgements

This work has been supported in part by the Grant-in-Aid for Scientific 
Research (3730) and Center-of-Excellence (COE) research (07CE2002) of the 
Ministry of Education, Science, and Culture of Japan. We would like to 
thank Bruce A. Peterson for many fruitful discussions.

\vspace{3cm}
\begin{center}
Table 1 \\
Sensitivity of the estimated age and metallicity to the change of the 
model parameters \\
\medskip
\begin{tabular}{lcc}
\hline\hline 
                         & $t$(std)=1.50 Gyr & [Fe/H](std)$=+0.96$ \\
\cline{2-3} 
\multicolumn{1}{c}{Changes} 
                         &   $\Delta t$      &    $\Delta$ [Fe/H]  \\
\hline
$[I($Fe$)/I($Mg$)]_\odot=3$ 
                $\to$  1.5/4  &  $+1.05/-0.13$  &  $+0.32/-0.10$  \\
SFR $\nu_{k=1}=7.6$ $\to$  6.7    &      $+0.08$    &     $+0.01$  \\
SFR $k=1$       $\to$   2$^a$ &      $+0.63$    &     $-0.01$     \\
IMF $x=1.35$    $\to$   1.1   &      $+1.15$    &     $+0.39$     \\
IMF $m_l=0.05$  $\to$   0.1   &      $+0.26$    &     $+0.08$     \\
SN Ia $A=0.055$ $\to$   0.08  &      $-0.14$    &     $+0.04$     \\
SN Ia $g$-form: $\gamma=0$ $\to$ 1$^b$ 
                              &      $+0.21$    &     $-0.04$     \\
\hline\hline
\end{tabular}
\vskip 0.2pc
\end{center}
\begin{flushleft}
\hspace*{0cm}$^a$ The coefficient $\nu_{k=2}$ is chosen so that 
the gas fraction at 0.5 Gyr should be the same as that from $\nu_{k=1}=7.6$ 
Gyr$^{-1}$. \\
\hspace*{0cm}$^b$ The power-law index 
$\gamma=1$ corresponds to a case 
that SN Ia progenitors of longer lifetime, which are possibly less 
massive, are given a higher weight like the IMF.  
\end{flushleft}

\begin{figure}[h]
\centerline{\psfig{figure=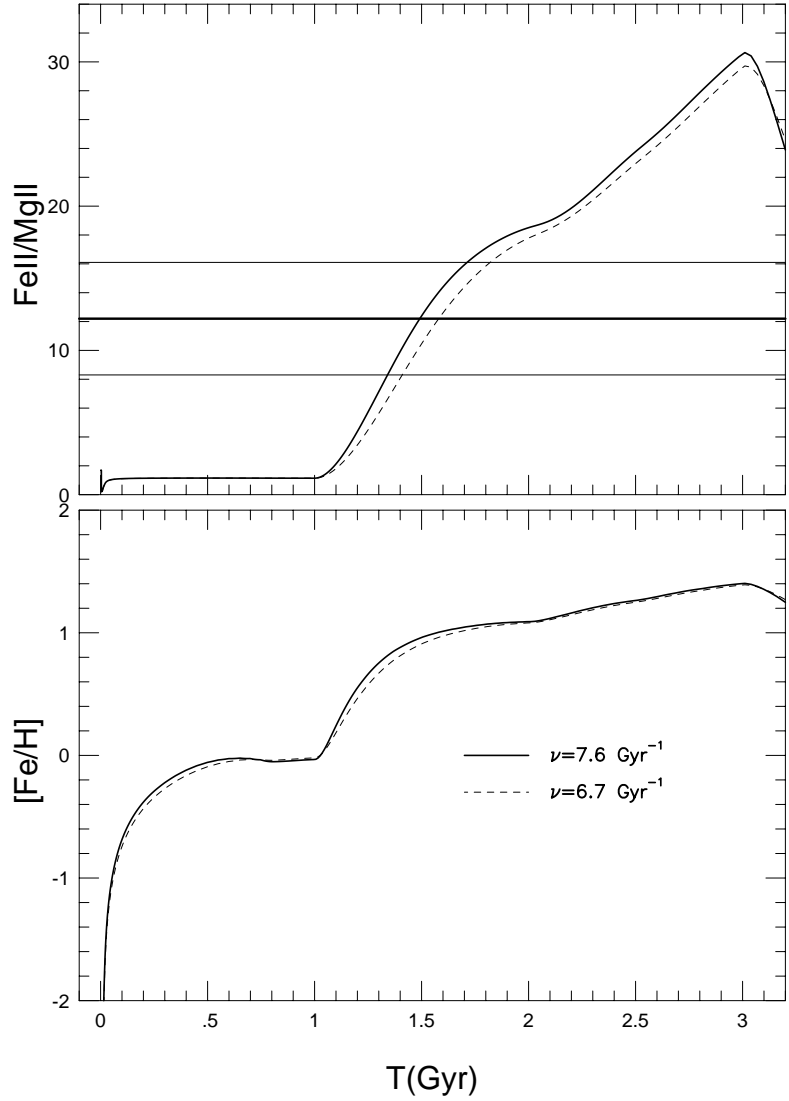,width=13.3cm}}
\caption[]{ Evolutionary behaviors of the Fe{\scriptsize
II}/Mg{\scriptsize II} flux ratio ({\it upper panel}) and the
logarithmic iron abundance [Fe/H] ({\it lower panel}) for theoretical
models calculated with different SFRs.  Shown are the standard model
with $\nu_{k=1}=7.6$ Gyr$^{-1}$ ({\it thick line}) and the lower SFR
model with $\nu_{k=1}=6.7$ Gyr$^{-1}$ ({\it dashed line}).  The
horizontal line is the value of the Fe{\scriptsize II}/Mg{\scriptsize
II} flux ratio measured for B1422+231, and the shaded area is the
uncertainty in this measurement. 
}
\end{figure}

\begin{figure}[h]
\centerline{\psfig{figure=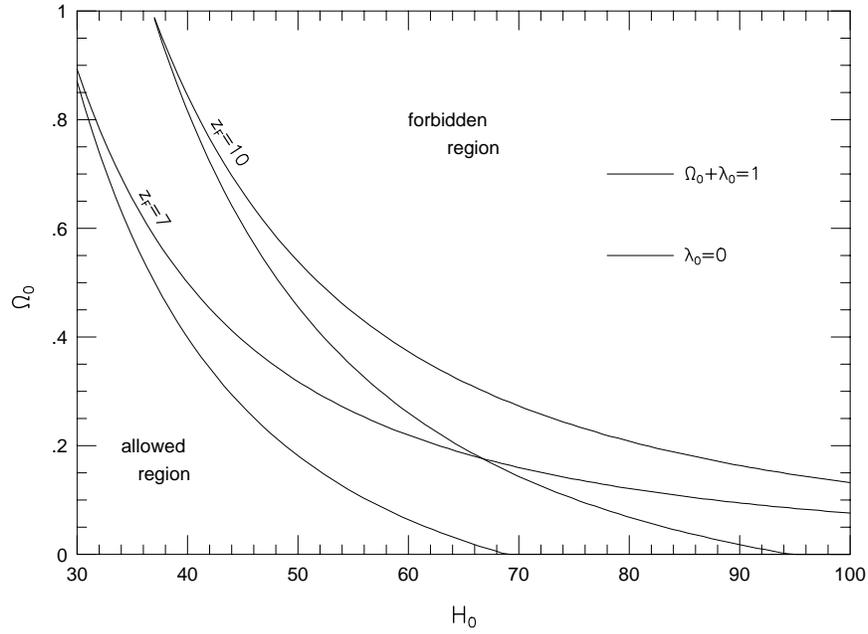,width=13.3cm}}
\caption[]{ 
Constraints on the cosmological parameters ($H_0$, $\Omega_0$, 
$\lambda_0$)
and the epoch of galaxy formation $z_F$.  Shown are the $\Omega_0$ versus
$H_0$ relations derived from the lower age bound of 1.3 Gyr for B1422+231 
at $z=3.62$.  Thin lines represent the cases with $z_F=7$ and 10 for a 
universe with $\lambda_0=0$, and thick lines are those for a flat universe 
with $\Omega_0+\lambda_0=1$.  The area below each line is the allowed 
reigion where the expansion age of the universe exceeds the lower age 
bound of 1.3 Gyr.  Note that current estimates of the Hubble constant are 
in the range of $H_0=60-80$ km\,s$^{-1}$\,Mpc$^{-1}$.
}
\end{figure}

\begin{figure}[h]
\centerline{\psfig{figure=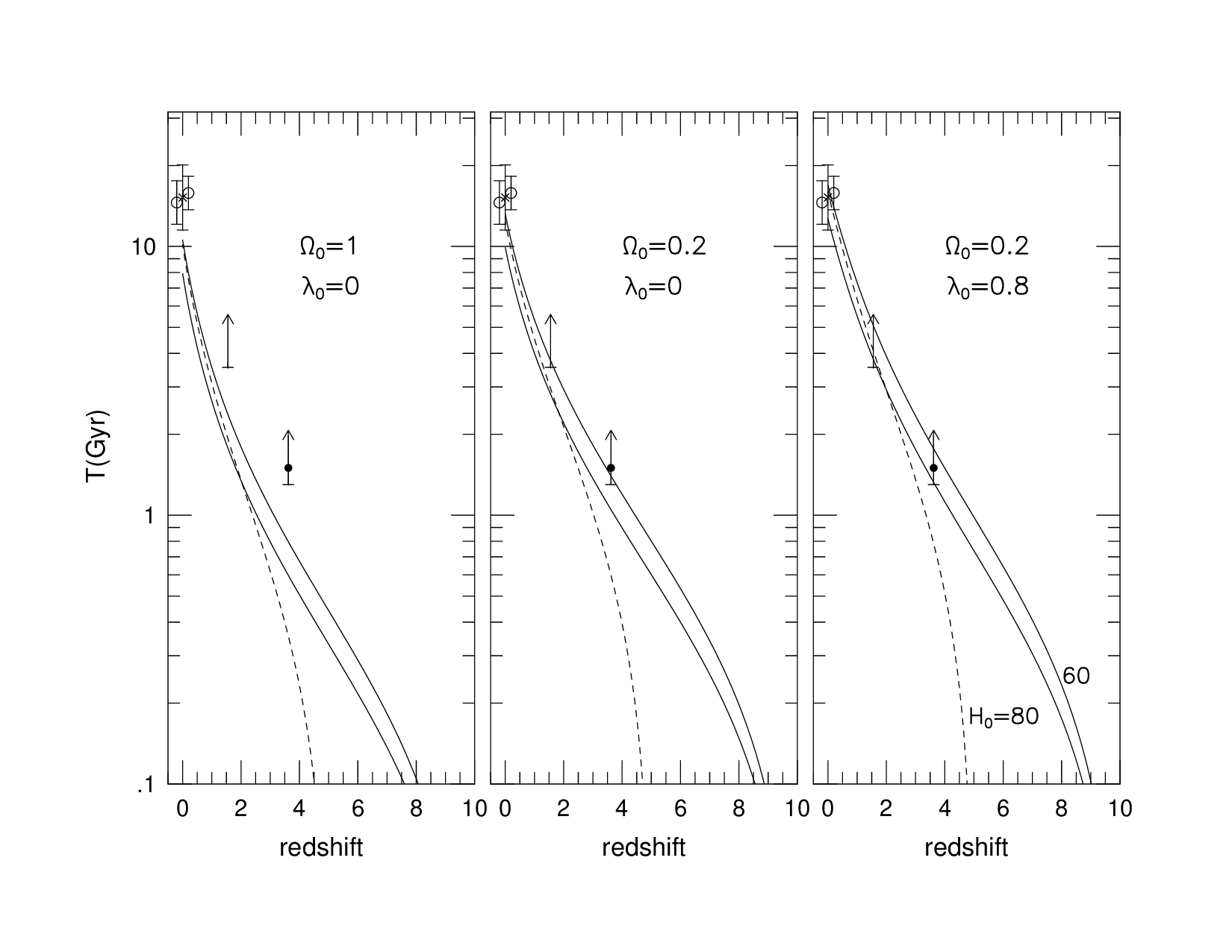,width=15cm}}
\caption[]{ 
The age--redshift relations predicted for the Einstein-de 
Sitter universe or a flat, matter-dominated universe with 
$(\Omega_0, \lambda_0)=(1,0)$ ({\it left panel}), an open universe with 
$(\Omega_0, \lambda_0)=(0.2,0)$ ({\it middle panel}), and a flat, 
$\Lambda$-dominated universe with $(\Omega_0, \lambda_0)=(0.8,0.2)$ 
({\it right panel}).  In each panel shown are the predictions for $z_F=10$ 
with $H_0$= 60 and 80 km\,s$^{-1}$\,Mpc$^{-1}$ ({\it solid lines}) and 
for $z_F=5$ and $H_0$= 60 km\,s$^{-1}$\,Mpc$^{-1}$ ({\it dashed line}).  
The age of 1.5 Gyr for B1422+231 estimated by the standard model of chemical
enrichment is indicated at $z=3.62$ by the filled circle together with its 
lower age bound of 1.3 Gyr.  As a summary of the existing age determinations
at $z\approx 0$, the ages of GCs (15.8$\pm$2.1 Gyr, Bolte \& Hogan 1995; 
14.56$\pm$2.49 Gyr, Chaboyer et al. 1996) are shown by the open circles and 
the age of the CS22892-052 star ($15.2\pm3.7$ Gyr, Cowan et al. 1997) by the 
cross.  The age of LBDS 53W091 ($\gtsim 3.5$ Gyr, Spinrad et al. 1997) is 
also indicated at $z=1.55$.
}
\end{figure}


\begin{thebibliography}{}  

\bibitem[]{}
Bolte, M., \& Hogan, C. J. 1995, Nature, 376, 399

\bibitem[]{} 
Carney, B. W. 1996, PASP, 108, 900

\bibitem[]{} 
Chaboyer, B., Demarque, P., Kernan, P. J., \& Krauss, L. M. 
1996, Science, 271, 957

\bibitem[]{} 
Collin-Souffrin, S., Hameury, J.-M., \& Joly, M. 
1988, A\&A, 205, 19

\bibitem[]{} 
Contini, T., Davoust, E., \& Consid\`{e}re, S. 1995, A\&A, 303, 440

\bibitem[]{} 
Cowan, J. J., McWilliam, A., Sneden, C., \& Burris, D. L. 
1997, ApJ, 480, 246

\bibitem[]{}
Elston, R., Thompson, K. L., \& Hill, G. J. 1994, Nature, 367, 250

\bibitem[]{}
Feast, M. W. \& Catchpole, R. W. 1997, MNRAS, 286, L1

\bibitem[]{}
Francis, P. J., Hewett, P. C., Foltz, C. B., Chaffee, F. H., 
Weymann, R. J., \& Morris, S. L. 1991, ApJ, 373, 465

\bibitem[]{}
Freedman, W. L. 1998, in the 18th Texas Symp., eds. A. Olinto, J. Frieman,
and D. Schramm (World Scientific Press), in press

\bibitem[]{}
Garnavich, P. M., et al. 1998, ApJ, 493, L53

\bibitem[]{}
Hamann, F., \& Ferland, G. 1993, ApJ, 418, 11

\bibitem[]{}
Hamann, F., \& Ferland, G. 1992, ApJ, 391, L53

\bibitem[]{} 
Kawara, K., Murayama, T., Taniguchi, Y. \& Arimoto, N. 
1996, ApJ, 470, L85

\bibitem[]{} 
Kennicutt Jr, R. C. 1996, Nature, 381, 555

\bibitem[]{} 
Kobayashi, C., Tsujimoto, T., Nomoto, K., Hachisu, I, \& Kato, M. 
1998, ApJ, 503, L155

\bibitem[]{} 
Low, C., \& Lynden-Bell, D. 1976, MNRAS, 176, 367

\bibitem[]{} 
Matteucci, F., \& Padovani, P. 1993, ApJ, 419, 485

\bibitem[]{}
Netzer, H., \& Wills, B. J. 1983, ApJ, 275, 445

\bibitem[]{}
Nomoto, K., Thielemann, F.-K., \& Wheeler, J. C. 1984, ApJ, 279, L23

\bibitem[]{}
Padovani, P., \& Matteucci, F. 1993, ApJ, 416, 26

\bibitem[]{}
Perlmutter, S., et al. 1998, Nature, 391, 51

\bibitem[]{}
Reid, I. N. 1997, AJ, 114, 161

\bibitem[]{} 
Renzini, A., \& Buzzoni, A. 1986, in The Spectral Evolution of Galaxies, 
ed. C. Chiosi \& A. Renzini (Dordrecht: Reidel), 195

\bibitem[]{}
Rieke, G. H., Loken, K., Rieke, M. J., \& Tamblyn, P. 1993, ApJ, 412, 99

\bibitem[]{} 
Sigut, T. A. A., \& Pradhan, A. K.  1998, ApJ, 499, L139

\bibitem[]{} 
Silk, J. 1977, ApJ, 214, 152

\bibitem[]{}
Spinrad, H., Dey, A., Stern, D., Dunlop, J., Peacock, J., Jimenez, R.,
\& Windhorst, R.  1997, ApJ, 484, 581

\bibitem[]{} 
Tinsley, B. M. 1980, Fundam.Cosmic Phys., 5, 287

\bibitem[]{} 
Truran, J. W. 1987, in the 13th Texas Symp. on Relativstic Astrophysics, 
ed. M. P. Ulmer (World Scientific: Singapore), 430

\bibitem[]{}
Tsujimoto, T., \& Yoshii, Y. 1998, in preparation

\bibitem[]{}
Tsujimoto, T., Miyamoto, M., \& Yoshii, Y. 1998, ApJ, 492, L79

\bibitem[]{}
Tsujimoto, T., Yoshii, Y., Nomoto, K., Matteucci, F., Thielemann, F.-K., \& 
Hashimoto, M. 1997, ApJ, 483, 228

\bibitem[]{}
Tsujimoto, T., Nomoto, K., Yoshii, Y., Hashimoto, M., Yanagida, S., \&
Thielemann, F.-K. 1995, MNRAS, 277, 945

\bibitem[]{}
Turner, E. L. 1991, AJ, 101, 5

\bibitem[]{}
Wills, B. J., Netzer, H., \& Wills, D. 1985, ApJ, 288, 94

\bibitem[]{}
Yoshii, Y., Tsujimoto, T., \& Nomoto, K. 1996, ApJ, 462, 266

\end{thebibliography}
\end{document}